\documentclass{ws-procs9x6}

\renewcommand\L{\mathbf{L}}

\newcommand\bbar[1]{\overline{#1}}

\newcommand{\ui}{\textrm{i}}
\newcommand{\ue}{\textrm{e}}

\newcommand{\ud}{\mathrm{d}}

\newcommand{\gz}{{\mathbb Z}}

\newcommand{\re}{\Re}

\newcommand{\vp}{\boldsymbol{\psi}}

\newcommand{\vz}{\mathbf{0}}

\newcommand{\rk}{\mathrm{rank}}

\newcommand{\A}{{\mathbb A}}
\newcommand{\B}{{\mathbb B}}

\newcommand{\bdm}{\begin{displaymath}}
\newcommand{\edm}{\end{displaymath}}
\newcommand{\beq}{\begin{equation}}
\newcommand{\eeq}{\end{equation}}
\newcommand{\beqa}{\begin{eqnarray}}
\newcommand{\eeqa}{\end{eqnarray}}

\newcommand{\cL}{\mathcal{L}}

\newcommand{\gB}{\mathcal{B}}

\newcommand{\V}{\mathcal{V}}

\newcommand{\csch}{\mathrm{csch}}

\begin{document}

\title{Vacuum energy, spectral determinant and heat kernel asymptotics of graph Laplacians with general vertex matching conditions}

\author{J. M. Harrison$^*$ and K. Kirsten$^\dag$}

\address{Department of Mathematics, Baylor University,\\
Waco, TX 76798, USA\\
$^*$E-mail: jon\_harrison@baylor.edu\\
$^\dag$E-mail: klaus\_kirsten@baylor.edu\\}

\begin{abstract}
We consider Laplace operators on metric graphs, networks of one-dimensional line segments (bonds), with matching conditions at the vertices that make the operator self-adjoint.  Such quantum graphs provide a simple model of quantum mechanics in a classically chaotic system with multiple scales corresponding to the lengths of the bonds.
For graph Laplacians we briefly report results for the spectral determinant, vacuum energy and heat kernel asymptotics of general graphs in terms of the vertex matching conditions.
\end{abstract}


\bodymatter
\section{Introduction}
Quantum graphs provide a simple, analytically tractable, model in which to investigate quantum phenomena.  This is of particular interest for the study of vacuum energy and related spectral quantities as the spectrum of a quantum graph generically incorporates the features associated with quantum mechanical systems that are classically chaotic\cite{p:GS:QG}.  In particular the eigenvalues of the Laplace operator on a graph are typically distributed as the eigenvalues of large random matrices where the appropriate random matrix ensemble for comparison is determined by the symmetries of the quantum system; this correspondence is known as the Bohigas-Giannoni-Schmit conjecture\cite{p:BGS:CCQS}.

Quantum graphs were introduced as a model of quantum chaos by Kottos and Smilansky\cite{p:KS:QCG}.  Graph models are also of current interest in many areas of mesoscopic physics like Anderson localization, photonic crystals, microelectronics, nanotechnology and the theory of wave guides; see Ref.~\refcite{p:K:GMWPTS} for a review. In this article we present results for the vacuum energy\cite{p:FKW:VERCFQSG,p:BHW:MAVEQG}, spectral determinant\cite{p:D:SDGGBC,p:D:SDSOG,p:F:DSOMG} and heat kernel asymptotics\cite{p:KPS:HKMGTF,p:BE:TFQGGSABC} of graph Laplacians.   The unified approach we adopt, based on the spectral zeta function of the graph, provides new results in each case.   The full description of the construction of zeta functions of quantum graphs will appear in Ref.~\refcite{p:HK:ZQG}.

\section{Graph Laplacian}
\begin{figure}[htb]
\begin{center}
\includegraphics[width=2.8cm]{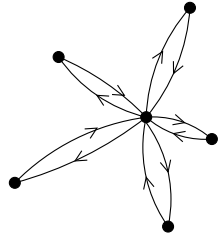}
\caption{A star (or hydra) graph.}\label{fig:star}
\end{center}
\end{figure}

Let $G = (\V,\gB)$ be a graph where $\V$ is the set of vertices and $\gB$ is the set of bonds (or edges); for an example see Fig. \ref{fig:star}.  Each bond $b\in \gB$ joins a pair of vertices
and for notational convenience we assume that bonds are directed with an initial and terminal vertex specified via the functions
$o:\gB\to\V$ and $t:\gB\to\V$ respectively.  We insist that the set $\gB$ is symmetric
so that $b \in \gB$ if and only if there is another bond
$\bbar{b} \in \gB$ the reversal of $b$ such that $o(b)=t(\bbar{b})$ and
$t(b)=o(\bbar{b})$.  The total number of bonds $B = |\gB|$ is therefore twice the physical number of connections in the network.
Each bond $b$ is associated with an interval $[0,L_b]$ so that the local coordinate $x_b=0$ at $o(b)$ and $x_b=L_b$ at $t(b)$.  We naturally require that $L_{\bbar{b}}=L_b$ so there is a single length associated with a physical connection in the network, consequently $x_{\bbar{b}}=L_b-x_b$.  The total length of the graph $G$ is denoted by $\cL=\frac{1}{2}\sum_{b=1}^B L_b$.

A function $\psi$ on $G$ is defined by the collection of functions
$\{\psi_b(x_b)\}_{b\in\gB}$ on the bonds of the graph.
We require $\psi_{b}(x_b) = \psi_{\bbar{b}}\left(L_b-x_b\right)$ in order that $\psi$ be physically meaningful.  The Hilbert space on the graph is
\begin{equation}\label{eq:Hilbert space}
    L^2(G) = \bigoplus_{b=1}^B L^2 \bigl( [0,L_b] \bigr) \ .
\end{equation}
To define a Laplacian on $G$ we consider the operator $-\frac{\ud^2}{\ud x_b^2}$ on the bonds with matching conditions between functions $\psi_b$ meeting at vertices of $G$ to make the graph Laplacian self-adjoint: for example Neumann like matching conditions require $\psi$ to be continuous at the vertices and
%
$\sum_{b|o(b)=v} \psi_b'(0)=0$
%
for all vertices $v\in \V$.
All such self-adjoint matching conditions were classified by Kostrykin and Schrader\cite{p:KS:KRQW} and we follow their scheme.  The matching conditions on the whole graph are specified by the pair of $B\times B$ matrices $\A,\B$ via
\begin{equation}\label{eq:matching conditions}
    \A \vp + \B \vp'=\vz \ ,
\end{equation}
where  $\vp=\big(\psi_1(0),\dots,\psi_B(0)\big)^T$ and $\vp'=\big(\psi'_1(0),\dots,\psi'_B(0)\big)^T$.
These matching conditions define the domain of a self-adjoint Laplace operator iff
$\rk (\A,\B)=B$ and $\A\B^\dag=\B\A^\dag$.

To find the graph spectrum we want to solve the Laplace eq. on the bonds
\begin{equation}
  \label{eq:Laplace_eq}
  -\frac{\ud^2}{\ud x_b^2} \psi_b(x_b) = k^2 \psi_b(x_b) \ .
\end{equation}
In the case of a star graph, Fig. \ref{fig:star}, solving Eq. (\ref{eq:Laplace_eq}) on the bonds and
substituting in the matching condition at the center one obtains the well known secular eq. of the star graph
%
$\sum_{b=1}^B \tan k L_b = 0$, whose
solutions $\{k_j |k_j\geqslant 0 \}$ are square roots of the eigenvalues of the Laplacian on the star.

If instead one substitutes solutions of (\ref{eq:Laplace_eq}) in the general form of matching conditions (\ref{eq:matching conditions}) we find the sequence $\{k_j \}$ are the solutions of a secular eq. $f(k)=0$ where,
\begin{equation}\label{eq:gen secular}
    f(k)=\det \left( \A
    + k \B
    \left( \begin{array}{cc}
    -\cot(k \L) & \csc(k \L)\\
    \csc(k \L)& - \cot( k \L)\\
    \end{array} \right)
    \right) \ ,
\end{equation}
and  $\L=\textrm{diag}\{L_1,\dots, L_B\}$.
This can be compared to the scattering matrix formulation introduced by Kottos and Smilansky\cite{p:KS:QCG,p:KS:POTSSQG}.

\section{Results}
The zeta function of the quantum graph can be formulated as a contour integral applying the argument principle to (\ref{eq:gen secular}) following the technique introduced in Refs.~\refcite{p:KM:FDCIM, p:KM:FDGSLP}, namely
\begin{equation}\label{eq:contour int}
    \zeta(s)=\sum_{j=0}^\infty \phantom{|}^{\prime} \, k_j^{-2s}=\frac{1}{2\pi \ui} \int_c z^{-2s} \frac{f'(z)}{f(z)}\,  \ud z = \frac{1}{2\pi \ui} \int_c z^{-2s} \frac \ud {\ud z}  \log f(z)\,  \ud z \ .
\end{equation}
The prime on the sum denotes the exclusion of zero modes if present.
Generically the poles of $f$ are the whole of the set $\{ m\pi/L_b | m\in \gz, b=1,\dots,B \}$ and this is the case we consider.\footnote{It is possible for particular choices of matching conditions $\A, \B$ and bond lengths $\{L_b\}$ that individual poles cancel in the determinant, however, a small perturbation of the bond lengths removes this.}  The contour $c$ is then chosen so as to include the zeros of $f$ on the positive real axis and exclude the poles. 
Making a contour transformation to an integral along the imaginary axis closed with a semicircular arc on the right we obtain an expression for $\zeta(s)$ valid in the strip $0<\re \, s < 1$,
\begin{eqnarray}\label{eq:gen zeta}
    \zeta(s)&=& \zeta_{R}(2s) \sum_{b=1}^{B} \Big( \frac{\pi}{L_b} \Big)^{-2s}
    +\frac{\sin \pi s}{\pi} \int_0^\infty t^{-2s} \frac{\ud}{\ud t} \log \hat{f}(t) \, \ud t \ ,\\
    \hat{f}(t)&=& \det \left( \A
    - t \B
    \left( \begin{array}{cc}
    \coth(t \L) & -\csch (t \L)\\
    -\csch (t \L)&  \coth(t \L)\\
    \end{array} \right)
    \right)  \ ,
\end{eqnarray}
where $\zeta_{R}$ is the Riemann zeta function.

Regularized formulations of the vacuum energy, and the spectral determinant as well as the heat kernel asymptotics are obtained directly from the zeta function.  The vacuum energy, formally $E=\frac{1}{2} \sum_{j=0}^{\infty} \phantom{|}^{\prime} k_j$ where $k_j^2$ is an eigenvalue of the Laplacian, is
defined as $E_c=\frac{1}{2} \zeta ( -1/2 )$.  Differentiating with respect to the $L_\beta$, the length of the bond $\beta$, we obtain the Casimir force on the bond,
\begin{equation}
    F_c^\beta =\frac{\pi}{24 L_\beta^2} + \frac{1}{\pi}\int_0^\infty \frac{\partial}{\partial L_\beta} \log \hat{f}(t) \, \ud t \ .
\end{equation}

The spectral determinant is formally the product of the eigenvalues $\prod_{j=0}^\infty \phantom{|}^{\prime} k^2_j$.  The zeta function representation (\ref{eq:gen zeta}) allows a direct evaluation of the regularized spectral determinant, as
\beq\label{eq:spectral det}
    {\det}' (-\triangle)=\exp\big( -\zeta'(0) \big)= \frac{2^B \hat{f}(0)}{c_N \prod_{b=1}^B L_b} \ .
\eeq
Here, $c_N$ denotes the first non-zero coefficient in the $t\to \infty$ asymptotic expansion of
\begin{equation}\label{eq:gen fhat infinity}
    \hat{f}(t) \sim
    \det \left( \A
    - t \B
    \right) = \det \B \, t^{2B} + c_{2B-1} t^{2B-1} + \dots + c_{1} t +\det \A \ .
\end{equation}
So $c_N=\det \B$ when $\det \B\ne 0$.  The factor of $c_N$ in (\ref{eq:spectral det}) is introduced by the analytic continuation of $\zeta(s)$ to $s=0$.

The heat kernel for a quasi one-dimensional system has an expansion for $t\to 0$ of the form,
$K(t)=\sum_{j=1}^\infty\ue^{-k_j^2 t}\sim \sum_{\ell =0,1/2,1,\dots}^{\infty} a_\ell t^{\ell-1/2}$.
The heat kernel coefficients are related to the zeta function\cite{seel68-10-288} by
$a_\ell = \textrm{Res} (\zeta(s) \Gamma(s) )|_{s=1/2-\ell}$.
Subtracting sufficiently many terms in the $t\to \infty$ expansion of $\hat{f}$, the zeta function (\ref{eq:gen zeta}) can be extended as far left of the imaginary axis as required and we obtain the full asymptotic expansion of the heat kernel of the general graph Laplacian,
\beq\label{heatasygen}
K(t) \sim \frac{{\cal L}} {\sqrt{4\pi t}} +\frac N 2  - \sum_{k=1,3/2,2,...} ^\infty \frac{b_{2k-1}}{\Gamma \left( k-\frac 1 2\right) }\, t^{k-1/2}\ .
\eeq
The coefficients $b_n$ are obtained from the expansion of $\log \hat{f}(t)$ at infinity,
\beq
\log \hat f (t) \sim  N \log t + \log c_N + \sum_{n=1} ^\infty \frac{b_n} {t^n} \ .
\eeq
For a given set of matching conditions defined by matrices $\A,\B$ the coefficients $b_n$ can be easily obtained with a computer algebra package.


\section*{Acknowledgments}
The authors would like to thank G Berkolaiko, JP Keating, P Kuchment, J Marklof, R Piziak and B Winn for helpful suggestions.
KK is supported by National Science Foundation grant PHY--0554849 and JMH is supported by National Science Foundation grant DMS--0604859.

\newpage
%
%
%

\end{document}